\documentclass[pop,fleqn]{w-art}
\usepackage{latexsym}
\usepackage{amsfonts}
\usepackage{amsmath}
\usepackage{times}
\usepackage{graphicx}

\begin{document}


\title{Progress on Holographic Three-Point Functions}

\author{Wolfgang M\"uck\footnote{e-mail: \textsf{mueck@na.infn.it}}}
\address{Dipartimento di Scienze Fisiche, Universit\`a di Napoli
  ``Federico II'' and\\
  I.N.F.N.\ --- Sezione di Napoli\\
  Via Cintia, 80126 Napoli, Italy}

\begin{abstract}
The recently developed gauge-invariant formalism for the treatment of
fluctuations in holographic renormalization group (RG) flows overcomes most
of the previously encountered technical difficulties. I summarize the
formalism and present its application to the GPPZ flow, where
scattering amplitudes between glueball states have been calculated and
a set of selection rules been found.
\end{abstract}

\maketitle

\newcommand{\ie}{\emph{i.e.,\ }}
\newcommand{\eg}{\emph{e.g.,\ }}

\newcommand{\rmd}{\mathrm{d}}
\newcommand{\e}[1]{\mathrm{e}^{#1}}

\newcommand{\tg}{\tilde{g}}
\newcommand{\tR}{\tilde{R}}
\newcommand{\tG}[2]{\tilde{\Gamma}^{#1}_{\;#2}}

\newcommand{\bp}{\bar{\phi}}
\newcommand{\vp}{\varphi}
\newcommand{\G}{\mathcal{G}}
\newcommand{\R}{\mathcal{R}}

\newcommand{\mfa}{\mathfrak{a}}
\newcommand{\mfb}{\mathfrak{b}}
\newcommand{\mfc}{\mathfrak{c}}
\newcommand{\mfd}{\mathfrak{d}}
\newcommand{\mfe}{\mathfrak{e}}

\newcommand{\ha}{\hat{\mfa}}
\newcommand{\ca}{\check{\mfa}}

\newcommand{\Op}{\mathcal{O}}
\newcommand{\vev}[1]{\left\langle{#1}\right\rangle}
\newcommand{\Green}{\mathfrak{G}}
\newcommand{\bprop}{\mathfrak{K}}

\newcommand{\Of}{\mathcal{O}(f)}
\newcommand{\Ofn}[1]{\mathcal{O}(f^{#1})}

\newcommand{\htt}{{h^{TT}}}
\newcommand{\tE}{\tilde{E}}

\newcommand{\K}{\mathcal{K}}


\section{Introduction}
\label{intro}
The gauge/gravity correspondence provides an excellent tool for
obtaining relevant information about supersymmetric Yang-Mills (SYM)
theories from the study of their dual supergravity backgrounds, which
are generated by stacks of $D$-branes. The most celebrated duality---known
as the AdS/CFT correspondence---relates the superconformal
$\mathcal{N}=4$ SYM theory in four dimensions (in the planar limit and
at large 't~Hooft coupling) with the type IIB supergravity on an
$AdS_5 \times S^5$ background \cite{Maldacena:1997re}. More generally,
it is a paradigm that the dynamics of supergravity on an (asymptotically)
anti-de Sitter space encodes the correlation functions of its dual
(deformed) conformal field theory. This concept has been made
quantitatively precise by the AdS/CFT correspondence formula
\cite{Gubser:1998bc, Witten:1998qj} and by holographic renormalization
(see \cite{Bianchi:2001kw, Martelli:2002sp, Papadimitriou:2004ap}
and references therein). 

Holographic renormalization has taught us that the most efficient way
to obtain field theory correlators is to look directly at \emph{exact
one-point functions}, \ie one-point functions of gauge-invariant
operators in the presence of sources, which, in principle, contain the
information of all higher-point functions. Let me outline the
holographic calculation taking, for simplicity, a scalar field as
example. In a generic, asymptotically AdS, bulk space-time of
dimension $d+1$, a scalar field obeying the field equations can be
written as the sum of two asymptotic series,
\begin{equation}
\label{scalar_generic}
\phi(x,r) = \e{-(d-\Delta)r} (1+\cdots) \hat{\phi}(x) +
            \e{-\Delta r} (1+\cdots) \check{\phi}(x)~.
\end{equation}
It is assumed that the coefficient $\Delta$, which is the conformal
dimension of the dual operator $\Op$ of $\phi$, is restricted by $d/2 <
\Delta \leq d$, so that $\Op$ is relevant or marginal, and the first
series in \eqref{scalar_generic} is the leading one. The variable $r$
is the ``radial'' variable of the bulk, with $r\to \infty$
representing the asymptotic region. The ellipses
stand for higher order terms in the series, and the coefficients
$\hat{\phi}$ and $\check{\phi}$ are called the \emph{source} and
\emph{response} functions, respectively. A priori, these functions are
independent in the asymptotic analysis of the (second order) equations
of motion, but imposing a regularity condition in the bulk interior
functionally fixes the response to the source. 

The main result of holographic renormalization is that the exact
one-point function $\vev{\Op}$ is essentially given by the response
function, 
\begin{equation}
\label{exact1pt}
\vev{\Op_\Delta} = (2\Delta-d) \check{\phi} +\text{local terms.}
\end{equation}
The local terms are typically scheme dependent and will not be
discussed here. Hence, in order to calculate field theory
three-point functions involving $\Op_\Delta$, one must calculate
$\check{\phi}$ to quadratic order in the field fluctuations and then
differentiate with respect to the sources corresponding to the other
two operators.

In this talk, I will consider a generic bulk system of scalars and
gravity, governed by an action of the form 
\begin{equation}
\label{action5d}
  S = \int \rmd^{d+1}x \sqrt{\tg} \left[ -\frac14 \tR +\frac12
  G_{ab}(\phi) \partial_\mu \phi^a \partial^\mu \phi^b +V(\phi) \right]~,
\end{equation}
where the potential, $V(\phi)$, is given in terms of a superpotential,
$W(\phi)$, by
\begin{equation}
\label{Vdef}
  V(\phi) = \frac12 G^{ab} W_a W_b -\frac{d}{d-1} W^2~,
\end{equation}
and the matrix $G^{ab}$ is the inverse of the $\sigma$-model metric
$G_{ab}$. The notation coincides with the one used in
\cite{Bianchi:2003ug}, \ie bulk quantities are adorned with a tilde,
and derivatives of the potentials with respect to fields are indicated
as subscripts, as in $W_a=\partial W/\partial\phi^a$.

The equations of motion stemming from the action \eqref{action5d}
allow for a particular class of solutions with $d$-dimensional
Poincar\'e invariance, which are called Poincar\'e domain walls, or
holographic RG flow backgrounds. These are governed by a system of
first order differential equations in terms of the radial
variable $r$ (see, \eg \cite{Freedman:1999gp}), which implies that the
background scalars, $\bp^a$, can have either a non-zero source, or a
non-zero response, but not both. In the first case, we speak of a 
\emph{deformation flow} generated by the insertion of the dual
operator, whereas, in the second case we speak
of a \emph{vev flow}, since, according to \eqref{exact1pt}, 
the dual operator acquires a non-zero vacuum expectation value.
In the common nomenclature, scalars with non-zero
background are called \emph{active}, while those with zero
background are called \emph{inert}.

When M.~Bianchi, M.~Prisco and I started working on the 
calculation of three-point functions in holographic RG flows, the
state-of-the-art were calculations of 
two-point functions \cite{DeWolfe:2000xi, Arutyunov:2000rq,
  Bianchi:2000sm, Muck:2001cy, Bianchi:2001de}, and some simple
three-point functions \cite{Bianchi:2003bd}. Technical difficulties,
which were known already from the linear analysis needed for the
two-point functions, forced us to look for a new and more systematic
approach to the treatment of fluctuations in holographic RG flow
backgrounds. In this talk, I will introduce the
gauge-invariant formalism, which we developed in
\cite{Bianchi:2003ug} using reparametrization invariance as the
guiding principle. This formalism elegantly overcomes the above 
mentioned difficulties and has been successfully applied to the calculation
of scattering amplitudes in the GPPZ flow \cite{Muck:2004qg}. 

The generalization of the gauge-invariant formalism to a generic bulk system
of the form \eqref{action5d} has been undertaken in an on-going
collaboration with M.~Berg and M.~Haack, in which we hope to learn
something about holography in non-asymptotically AdS RG flow
backgrounds. In fact, supergravity-type
actions of the form \eqref{action5d} and holographic RG flow
backgrounds appear also in connection with a whole list of ``famous''
supergravity duals of SYM theories, \eg the Klebanov-Strassler
\cite{Klebanov:2000hb} and Maldacena-Nu{\~n}ez (MN)
\cite{Maldacena:2000yy} solutions.

\section{Gauge invariant formalism}
\label{gaugeinv}

\subsection{The $\sigma$-model covariant field expansion}
\label{covnot}
In the generic bulk system, which is described by the action
\eqref{action5d}, reparametrization invariance appears at two distinct
places, namely, in the geometries of the bulk space and of the
$\sigma$-model.
As usual in gravity, invariance under space-time diffeomorphisms comes
at the price of introducing redundant variables to the metric 
degrees of freedom. Usually, this is taken care of by gauge fixing,
but our approach will be different. For now, I shall keep all metric degrees
of freedom and describe later how to isolate the physical ones. The
Poincar\'e-invariant form of the holographic RG flow backgrounds then
suggests to use the ADM formalism to parametrize the metric degrees of 
freedom, with $g_{ij}$, $n^i$ and $n$ being the induced metric on the
time-slice hypersurfaces, the shift vector and the lapse function,
respectively. As they are tensors (of rank 2, 1 and 0) on the
time-slice hypersurfaces, we make sure to follow our guiding 
principle. The expansion of the metric around the background 
configuration is done by writing
\begin{equation}
\label{expandmetric}
  g_{ij} = \e{2A(r)} \left( \eta_{ij} + h_{ij} \right)~,\quad
  n_i    = \nu_i~,\quad
  n      = 1+ \nu~,
\end{equation}
where $h_{ij}$, $\nu_i$ and $\nu$ denote small
fluctuations. Henceforth, I shall adopt the notation that the indices
of the fluctuations, as well as of the derivatives $\partial_i$, 
are raised and lowered using the flat (Minkowski/Euclidean) metric.
It turns out to be useful to decompose the metric fluctuations into
\begin{equation}
\label{hsplit}
 h^i_j = \htt^i_j 
  + \partial^i \epsilon_j +\partial_j \epsilon^i
  + \frac{\partial^i \partial_j}{\Box} H + \frac1{d-1} \delta^i_j h~,
\end{equation}
where $\htt^i_j$ denotes the traceless transversal part, and $\epsilon^i$
is a transversal vector.

The space of fields $\phi^a$ has its own geometry,
characterized by the $\sigma$-model metric $G_{ab}$, which is assumed
to be invertible, the inverse being called $G^{ab}$. Hence, one can 
straightforwardly define the $\sigma$-model connection, $\G^a_{bc}$, the
Riemann curvature tensor, $\R^a_{bcd}$, and covariant field
derivatives, denoted by $D_a$ or by placing a bar $|$ before the
field index. Moreover, as the background is
$r$-dependent, it is useful to introduce also a ``background
covariant'' derivative, $D_r$, acting on tensors in field space, such
that $G_{ab}$, evaluated on the background, is covariantly constant.

In order to exploit this notation for the fluctuation equations, it is
necessary to perform the expansion of the scalar fields in a
$\sigma$-model covariant fashion. As is well known, such an expansion
is provided by the \emph{exponential map}, whose generator will be
called $\vp^\mu$, which, geometrically, represents the tangent vector
of the geodesic connecting a background point $\bp$ with the point
$\phi$. In practice, calculations are simpler when carried out in
Riemann normal coordinates \cite{Petrov}.

\subsection{Gauge transformations and invariants}
\label{invars}
Our main argument \cite{Bianchi:2003ug}, which aims at obtaining the
equations of motion in an explicitly gauge-invariant form, starts by
considering the effect of diffeomorphisms on the fluctuation
fields. It is well known that a diffeomorphism of the form 
$  x^\mu\to x'{}^\mu=x^\mu- \xi^\mu(x)$,
where $\xi^\mu$ is infinitesimal, acts as a gauge transformation on the
fluctuation fields, when the background is unchanged. It turns out,
though, that one can, order by order, form combinations of the
fluctuation fields, which are gauge-invariant. To describe this, it is
easiest to proceed in a symbolic fashion. The fluctuation fields shall
be classified into two sets, $X=\{h,H,\epsilon^i\}$ and $Y=\{\varphi^a,
\nu, \nu^i, \htt^i_j\}$, and the gauge-invariant fields shall be
called $I=\{\mfa^a,\mfb,\mfc,\mfd^i,\mfe^i_j\}$, where $\mfd^i$ and
$\mfe^i_j$ are transversal and traceless transversal, respectively. 
The symbols $X$, $Y$, and $I$ shall also be used to denote the members
of the corresponding sets. Solving the definitions of the
gauge-invariant variables for $Y$ yields relations of the form 
\begin{equation}
\label{Yeq}
  Y = I + y(X) + \alpha(X,X) +\beta(X,I) +\Ofn{3}~,
\end{equation}
where $y(X)$ is a linear function, quadratic terms have been
included in the form of bi-linear 
functions $\alpha$ and $\beta$, and $\Ofn{n}$ denotes terms at least of order
$n$ in the fluctuations. Terms of the form $\gamma(I,I)$ do not
appear, as they can be absorbed into $I$. Hence, there is a one-to-one
correspondence between the fields $Y$ and the gauge-invariant
variables $I$. In contrast, the fields $X$ parametrize the (unphysical)
gauge degrees of freedom, as can be seen by considering the
transformation law of $h^i_j$. The explicit relations are of the form 
\begin{equation}
\label{xi_delX}
  \xi^\mu = z^\mu(\delta X) +\Ofn{2} = \delta z^\mu(X)
  +\Ofn{2}~,
\end{equation}
where $z^\mu(X)$ are linear functionals. 
Eqs.\ \eqref{Yeq} and \eqref{xi_delX} will play an essential role in
finding the gauge-invariant form of the equations of motion.

\subsection{Einstein's equations and gauge invariance}
\label{Einstein_inv}
It is our aim to re-write the equations of motion in a gauge-invariant
fashion. ``Gauge-invariant'' here means that the final equations
should contain only the fields $I$ and make no reference to $X$ and
$Y$. Our guiding principle tells us that this should be possible,
because the physical dynamics does not depend on the gauge. 

For brevity, we shall consider Einstein's equations, symbolically
written as $E_{\mu\nu} = 0$, but it is clear that the same arguments
also hold for the equations of motion for the scalar fields. 

Expanding $E_{\mu\nu}$ by brute force in the fields $X$ and $Y$, and
then substituting \eqref{Yeq} for $Y$ yields an expression in terms of
$I$ and $X$ of the form
\begin{equation}
\label{E_2}
  E_{\mu\nu} = E^{(1)1}_{\mu\nu}(X) + E^{(1)2}_{\mu\nu}(I)
              + E^{(2)1}_{\mu\nu}(X,X) + E^{(2)2}_{\mu\nu}(X,I) 
	      + E^{(2)3}_{\mu\nu}(I,I) +\Ofn{3}~.
\end{equation}
The background equation is satisfied identically, and the single terms
are linear or bi-linear functions of their arguments. 

Considering explicitly the gauge transformation of \eqref{E_2} and
comparing to what one expects from the general transformation law of
second-rank tensors under diffeomorphisms [using \eqref{xi_delX}], one finds
\begin{equation}
\begin{split}
  E^{(1)1}_{\mu\nu}(X) &= 0~,\qquad
  E^{(2)1}_{\mu\nu}(X,X) =0~,\\
  E^{(2)2}_{\mu\nu}(X,I) &=   
                  [\partial_\mu z^\lambda(X)] E^{(1)2}_{\lambda\nu}(I)
		 +[\partial_\nu z^\lambda(X)] E^{(1)2}_{\mu\lambda}(I) 
                 +z^\lambda(X) \partial_\lambda E^{(1)2}_{\mu\nu}(I)~.
\end{split}
\end{equation}
Thus, the gauge dependent terms, which appear in the brute force
expansion of $E_{\mu\nu}$ at second order, contain the first
order equation, and, in an order by order analysis, can be
consistently dropped. This argument generalizes recursively to higher
orders. One will find that the gauge dependent terms of any given
order can be consistently dropped, because they contain the equation
of motion at lower orders.

Happily, we have arrived at an equation of motion, which is written in
terms of the gauge-invariant variables $I$ only, and which, therefore,
is explicitly gauge-invariant. Thus, we have removed the unphysical
degrees of freedom without an explicit gauge fixing. The
gauge-invariant equations of motion are simply found by applying the
following substitution rules,
\begin{equation}
\label{field_subs}
   \varphi^a \to \mfa^a~,\quad
          \nu       \to \mfb~,\quad
          \nu^i     \to \mfd^i + \frac{\partial^i}{\Box} \mfc~,\quad
          h^i_j     \to \mfe^i_j~. 
\end{equation}
Since $\mfe^i_j$ is traceless and transversal, the calculational
simplifications stemming from \eqref{field_subs} are
considerable.

\subsection{Equations of motion}
\label{fieldeq}
The equations of motion that follow from the action \eqref{action5d}
are the equation for the scalar fields and Einstein's equations, the
latter being conveniently split into the normal components, $E_{rr}$,
the mixed components, $E_{ir}$, and the tangential components,
$E_{ij}$. After using the substitution rules \eqref{field_subs} one
finds that the scalars $\mfa^a$ physically couple to $\mfb$ and $\mfc$
at the linearized level, but, in contrast to the experience in the
literature, in the gauge-invariant formalism the components $E_{rr}$
and $E_{ir}$ can be solved algebraically for $\mfb$ and $\mfc$, so
that substituting them into the scalar equation of motion yields 
the compact expression
\begin{equation}
\label{eqphi}
  \left[ \left( \delta^a_b D_r +W^a_{\;\;|b} -\frac{W^aW_b}{W}
         -\frac{2d}{d-1} W \delta^a_b \right) 
         \left( \delta^b_c D_r -W^b_{\;\;|c} +\frac{W^bW_c}{W} \right)
  +\delta^a_c \e{-2A} \Box \right] \mfa^c =\tilde{J}^a~,
\end{equation}
where $\tilde{J}^a$ denotes quadratic interaction terms. 
The field $\mfd^i$ is suppressed at linear order, which is expected
from the boundary Ward identity, so that the only independent physical
degrees of freedom of the metric fluctuations are $\mfe^i_j$. Their
equation of motion is found from $E_{ij}$ and reads
\begin{equation}
\label{eqe}
  \left( \partial_r^2 -\frac{2d}{d-1} W \partial_r +\e{-2A} \Box
  \right) \mfe^i_j = J^i_j~.
\end{equation}
The quadratic interaction terms $\tilde{J}^a$ and $J^i_j$ can be found
in \cite{Bianchi:2003ug, Muck:2004qg} and depend on $\mfa^a$ and
$\mfe^i_j$.

Equation \eqref{eqphi} is the main result of the gauge-invariant
formalism. It governs
the dynamics of scalar fluctuations around Poincar\'e domain walls in
the most general case. Being a system of second order differential
equations, one can use the standard Green's function method
to treat the interactions perturbatively.

\section{Applications}
\subsection{Three-point functions in holographic RG flows}
The application of \eqref{eqphi} and \eqref{eqe} to three-point
functions in holographic RG flows is rather
straightforward. First, since both equations of motion are second
order differential equations ($\Box$ is replaced by $-k^2$ after
Fourier transforming into momentum space), it is easy to write down
their formal, non-linear, solutions. For example, 
\begin{equation}
\label{asol}
  \mfa^a(z) = \int \rmd^d y\, K^a_{a'}(z,y) \ha^{a'}(y) 
  + \int \rmd^{d+1} z'\, \sqrt{\tg(z')} \mathfrak{G}^a_{a'}(z,z')
  \tilde{J}^{a'}(z')~,
\end{equation}
where $K^a_{a'}(z,y)$ and $\mathfrak{G}^a_{a'}(z,z')$ denote the
bulk-to-boundary propagator and the bulk Green's function,
respectively. Moreover, $\ha^a$ are the prescribed sources for
the scalar fields. Substituting the free solutions into $\tilde{J}^a$
in the second term yields the solution to quadratic order. By determining
the asymptotic behaviour of $\mfa^a$ from \eqref{asol} one then finds
the response functions $\ca^a$ to quadratic order in the sources
$\ha^a$ and $\hat{\mfe}^i_j$. Hence, after differentiating twice with
respect to the sources and 
using \eqref{exact1pt} one finds the non-local terms of the
three-point functions. Generically, they are of the form
\begin{equation}
\label{3pt}
\vev{\Psi_1 \Psi_2 \Psi_3} = -\delta (p_1+p_2+p_3) \int \rmd r\, \e{dA}
\mathcal{X}_{123} K_1 K_2 K_3~,
\end{equation}
where $K_i$ ($i=1,2,3$) are the bulk-to-boundary propagators of the
fields dual to the operators $\Psi_i$, and the operator
$\mathcal{X}_{123}$ is easily read off from the interaction terms
$\tilde{J}^a$. To be more precise, a field re-definition to remove
terms in $\tilde{J}^a$ that have two $r$-derivatives, and an
integration by parts in \eqref{3pt} might be necessary in order to
achieve explicit bose symmetry of the correlators
\cite{Bianchi:2003ug, Muck:2004qg}.

\subsection{Scattering amplitudes in the GPPZ flow}
The GPPZ flow \cite{Girardello:1999bd} is the gravity dual of a
deformation of $\mathcal{N}=4$ SYM theory by the insertion of the
$\Delta=3$ operator $\Op=\delta_{AB} \mathrm{Tr} (X^A X^B)$, where
$A,B=1,2,3$, and $X^A$ are three out of the six scalar fields of
$\mathcal{N}=4$ SYM theory. It is $\mathcal{N}=1$ supersymmetric and
has qualitative features in common with pure $\mathcal{N}=1$ SYM
theory, such as confinement, but not the gluino condensate. The bulk
dynamics is governed by an action of the type \eqref{action5d}, with
one active scalar $\phi$ (dual to the inserted operator), one inert
scalar $\sigma$, which is dual to a scalar operator $\Sigma$ in the
gluino-bilinear multiplet $\mathrm{Tr}(W_\alpha W^\alpha)$, and, as
usual, the metric fluctuations are dual to the energy momentum
tensor, $T^i_j$. 

The holographic calculation of the two-point functions of the
operators $\Op$, $\Sigma$ and $T^i_j$ \cite{Girardello:1999bd,
  DeWolfe:2000xi, Arutyunov:2000rq, Bianchi:2000sm, Muck:2001cy,
  Bianchi:2001de} 
has revealed discrete, but infinite, spectra of states labelled by a
positive integer $k$, which are interpreted as glueballs of spin zero
($\Op_k$ and $\Sigma_k$) and spin two ($T_k$). The mass values of the
glueballs all summarized in the second column of table~1.

The start of our collaboration was motivated mostly by the physical 
question to calculate three-particle scattering amplitudes between the
above glueball states. In order to do this from holography, one must
calculate all three-point functions of the operators $\Op$, $\Sigma$
and $T^i_j$ from the bulk dynamics, and then amputate the external
legs in a standard field theory fashion. The
gauge-invariant formalism presented in this talk is, in our
opinion, the most suitable and elegant approach. As published
elsewhere \cite{Muck:2004qg}, we have calculated all ten independent
three-point functions of the above operators, the final results being given
in terms of bulk integrals of the form \eqref{3pt}. For the GPPZ flow,
the bulk-to-boundary propagators are hypergeometric functions, and one
cannot hope to perform the integrals explicitly. However, putting the
external momenta on-shell, each hypergeometric function reduces to a
product of the corresponding on-shell mass pole, which is amputated for the
scattering amplitudes, and a Jacobi polynomial. Hence, the bulk integrals for
the scattering amplitudes are elementary. Our analysis has revealed
that, surprisingly, most of the kinematically allowed scattering
amplitudes vanish. These selection rules have a deeper origin in
certain orthogonality relations between the Jacobi
polynomials. Table~1 summarizes the selection rules for the (kinematically
and dynamically) allowed glueball decay processes.

\begin{vchtable}[tb]
\vchcaption{Summary of mass spectra and allowed glueball decay
  processes. $L$ is the AdS length scale.} 
\begin{tabular}{|c|c|ll|}
\hline
glueball & $m^2L^2$ & \multicolumn{2}{c|}{decay channels}\\
\hline
               &          & $\to \Sigma_k + \Sigma_1$& $(k>1)$    \\
$\mathcal{O}_k$& $4k(k+1)$& $\to \Sigma_i + \Sigma_j$& $(i+j=k)$  \\
               &          & $\to \mathcal{O}_i + T_j$& $(i+j=k-1)$\\
\hline
           &              & $\to \Sigma_i +T_j$& $(i+j=k-1)$ \\
$\Sigma_k$ & $4(k-1)(k+2)$& $\to \Sigma_1 + \mathcal{O}_{k-1}$& \\
           &              & $\to \Sigma_1 + T_{k-1}$& \\
\hline
      &            & $\to \Sigma_i + \Sigma_j$ & $(i+j=k)$ \\
$T_k$ & $4(k+1)^2$ & $\to \Sigma_k + \Sigma_1$ & \\
      &  & $\to \mathcal{O}_i + \mathcal{O}_j$ & $(i+j=k)$ \\
\hline
\end{tabular}
\end{vchtable}

Although the GPPZ flow does not correctly capture the IR dynamics of the
$\mathcal{N}=1$ supersymmetric field theory (for example, the gluino
condensate is missing) we believe that the selection rules, which we
have found, are insensitive to the inclusion of further
non-perturbative effects, or, at least, that the effects of these
corrections are very weak. Hence, a comparison with field theory data,
\eg from lattice simulations of $\mathcal{N}=1$ SYM theory, would be
very interesting. 


\begin{acknowledgement}
I would like to thank my collaborators Marcus Berg, Massimo Bianchi,
Michael Haack and Maurizio Prisco for many inspiring discussions.
 
Financial support from the European Commission's RTN Programme,
contract HPRN-CT-2000-00131, from the Italian ministry of education
and research (MIUR), projects 2001-1025492 and 2003-023852, and from
INFN is gratefully acknowledged. 
\end{acknowledgement}


\end{document}